\documentclass[12pt]{article} 

\newcommand{\be}{\begin{equation}}
\newcommand{\ee}{\end{equation}}
\newcommand{\bea}{\begin{eqnarray}}
\newcommand{\eea}{\end{eqnarray}}

\newcommand{\gs}{\ensuremath{g_s}} 
\newcommand{\ls}{\ensuremath{l_s}} 
\newcommand{\lP}{\ensuremath{l_P}} 

\def\eps{\ensuremath{\epsilon}}
\def\half{\ensuremath{{1\over 2}}}

\def\p{\partial}

\def\para{{\scriptscriptstyle ||}}

\def\expec#1{\langle #1 \rangle}

\newcommand{\cF}{{\mathcal{F}}}

\newcommand{\cH}{{\mathcal{H}}}

\newcommand{\cN}{{\mathcal{N}}}

\newcommand{\Gs}{\ensuremath{G_s}} 
\newcommand{\Go}{\ensuremath{G_o^2}} 
\newcommand{\ape}{\ensuremath{\alpha'_{e}}} 
\newcommand{\Ls}{\ensuremath{L_s}} 
\newcommand{\LP}{\ensuremath{L_P}} 

\newcommand{\tbeta}{\ensuremath{\tilde{\beta}}}
\newcommand{\tgamma}{\ensuremath{\tilde{\gamma}}}

\newcommand{\tN}{\ensuremath{\tilde{N}}}

\newcommand{\Sng}{\ensuremath{S}}

\begin{document}

\begin{titlepage}

\begin{flushright}
ICN-UNAM-02/01\\
hep-th/0201140
\end{flushright}

\vspace{1cm}

\begin{center}
{\LARGE\bf A Membrane Action for OM Theory
}

\end{center}
\vspace{3mm}

\begin{center}

{\large J.~Antonio Garc\'{\i}a,
Alberto G\"uijosa,
and
J.~David Vergara}

\vspace{5mm}

Departamento de F\'{\i}sica de Altas Energ\'{\i}as \\
\vspace{1mm}
Instituto de Ciencias Nucleares, UNAM \\
\vspace{1mm}
Apdo. Postal 70-543, M\'exico, D.F. 04510

\vspace{5mm}

{\tt
garcia, alberto, vergara@nuclecu.unam.mx
}

\end{center}

\vspace{5mm}

\begin{center}
{\large \bf Abstract}
\end{center}
\noindent

Through direct examination of the effect of the OM limit on the M2-brane 
worldvolume action, we derive a membrane action for OM theory, and more 
generally, for the eleven-dimensional M-theoretic construct known as 
Galilean or Wrapped M2-brane (WM2) theory, which contains OM theory as a special 
class of states. In the static gauge,
the action in question implies a discrete 
spectrum for the closed membrane of WM2 theory, which under double 
dimensional reduction is shown to reproduce the known NCOS/Wound closed 
string spectrum. We examine as well open membranes ending on each of the 
three types of M5-branes in WM2 theory (OM theory arising from the 
`longitudinal' type), and show that the `fully transverse' fivebrane is 
tensionless. As a prelude to the membrane, we also study the case of the 
string, where we likewise obtain a reparametrization-invariant action, 
and make contact with previous work.

\vfill
\begin{flushleft}
January 2002
\end{flushleft}
\end{titlepage}
\newpage

\section{Introduction}
\label{introsec}

In recent times interesting limits of string/M theory have been 
discovered which give rise to decoupled open brane theories exhibiting 
some form of noncommutativity. Best understood among these are the 
$(p+1)$-dimensional Noncommutative Open String (NCOS) theories 
\cite{sst2,ncos}, defined as a low-energy limit of a stack of 
D$p$-branes in the presence of a near-critical worldvolume electric 
field (or, equivalently, a near-critical bulk Kalb-Ramond field 
$B_{01}$). While decoupling the usual closed string modes, the limit in 
question remarkably manages to retain the whole tower of open string 
excitations; the result is a non-gravitational open string theory which 
displays noncommutativity between space and time \cite{stnc}.

Shortly after the formulation of NCOS theories, generalizations based on 
other types of open branes were found \cite{om,bbss2,harmark2,odp,opq}. 
Foremost among these is Open Membrane (OM) theory \cite{om,bbss2}, 
defined as a low-energy limit of a stack of M5-branes with a 
near-critical worldvolume `electric' field strength (or, equivalently, 
in a near-critical bulk gauge field $C_{012}$). The limit yields a 
$(5+1)$-dimensional theory decoupled from gravity, which reduces to the 
$(2,0)$ superconformal theory at low energies. OM theory is expected to 
admit a description in terms of open M2-branes ending on the fivebranes, 
and to possess a generalized form of noncommutativity. This M-theoretic 
structure plays a role analogous to that of M theory itself: OM theory 
underlies and unifies all of the noncommutative theories which originate 
from string theory, be they of the open brane 
\cite{sst2,ncos,om,harmark2,odp,opq} or of the purely field-theoretic 
\cite{presw,sw} type.

The nature of NCOS theories came to be better understood through the 
work of Klebanov and Maldacena \cite{km}, who discovered that, upon 
compactification of the electric field direction, the NCOS spectrum 
contains not only open strings but also closed strings with strictly 
positive winding number. After that, it was shown in \cite{wound,go} that
the NCOS limit may be taken with or without 
D-branes, and consequently defines a $D$-dimensional string theory 
($D=10$ for the superstring).
In more detail: starting with any of the conventional string theories, 
one can single out a spatial direction (we will take it to be $x^{1}$, 
and refer to it as the longitudinal direction), compactify it on a 
circle of radius $R$, and consider an $\eps\to 0$ limit where the 
coupling constant, string length, (closed string) metric, and 
Kalb-Ramond field scale as
\be \label{strlim}
\gs={\Gs\over\sqrt{\eps}},\quad
\ls=\Ls\sqrt{\eps}, \quad
g_{\mu\nu}=(-1,1,\eps,\eps,\ldots), \quad
B_{01}=1-\lambda\eps,
\ee
with $\Gs,\Ls,R,\lambda$ fixed. This procedure yields a consistent 
$D$-dimensional theory, known as Wound \cite{wound} or Non-relativistic 
\cite{go} string theory, in which all objects carry strictly positive 
F-string winding along the longitudinal direction and are essentially 
non-relativistic. The parameters $\Gs$ and $\Ls$ introduced in 
(\ref{strlim}) are the effective coupling constant and string length
of the theory, and $\lambda$ is essentially a free (and for the most 
part physically irrelevant) parameter.\footnote{See \cite{wound,newt} 
for further characterization of these parameters and their relation to 
the NCOS parameters $\Go,\ape$.}
Since (\ref{strlim}) is the NCOS limit, NCOS theory evidently 
corresponds to the class of states in Wound string theory which contain 
D-branes extended along the longitudinal direction (with the theory on 
the branes decoupling from the bulk as $R\to\infty$). Other states in 
Wound string theory contain closed strings, transverse D-branes, and 
NS5-branes \cite{wound,go,br}.\footnote{An intriguing fact, with 
implications that remain to be determined, is that the transverse 
NS5-branes of Wound string theory are tensionless \cite{talk}. In 
Section \ref{omemsec} we will discuss these objects from an 
eleven-dimensional perspective.}
Gravity also turns out to be present, but in a vastly simplified form: 
it is Newtonian when the theory is formulated on a flat background 
\cite{go,newt}, and `asymptotically Newtonian' in a more general 
background \cite{newt,talk,sahakian2}.

Given the relation between NCOS and OM theory, the embedding of NCOS 
into Wound string theory implies an analogous embedding for the OM case. 
Indeed, Wound IIA string theory can be lifted to eleven dimensions to 
obtain what is known as Wrapped \cite{wound} or Galilean \cite{go} 
M2-brane theory, an M-theoretic construct which contains OM theory as a 
special class of states \cite{wound}.
Wrapped M2-brane (WM2) theory is defined as M theory compactified on a 
rectangular two-torus\footnote{One may also consider more general 
compactifications \cite{go,newt}.} with radii $R_1,R_2$, in an $\eps\to 
0$ limit where the Planck length, metric, and three-form gauge field 
scale as
\be \label{memlim}
\lP=\LP\eps^{1/3}, \quad
g_{MN}=(-1,1,1,\eps,\eps,\ldots), \quad
C_{012}=1-\lambda\eps,
\ee
with $\LP,R_1,R_2,\lambda$ fixed. All objects in this eleven-dimensional 
theory carry strictly positive M2 wrapping number on the $1$-$2$ torus, and 
are in effect non-relativistic. OM theory corresponds to those states of 
WM2 theory which contain M5-branes extended along the `longitudinal' 
directions $1$-$2$. WM2 theory contains in addition (partially or fully) 
transverse M5-branes,
closed M2-branes, and Newtonian gravity \cite{wound,go,newt}, and 
includes all Wound string and Wrapped brane theories \cite{wound,go} 
(and consequently all noncommutative open brane theories) as special 
limits. It is clearly desirable to increase our knowledge about this 
rich theoretical structure, which constitutes a simplified model of M 
theory.

OM theory is of course the most interesting subsector of WM2 theory. To 
date, information about it has been gathered either through direct 
examination of the effect of its defining limit, or by exploiting its 
connection to better understood theories.
Attempts have been made to understand its geometry 
\cite{bbss1,bbss2,ommetric} and noncommutative structure 
\cite{bbss1,bbss2,ncmem}, and its dual supergravity formulation has been 
scrutinized \cite{omsugra} (other work may  be found in \cite{otherom}).
In the NCOS case, the properties of the theory have for the most part 
been deduced in a similar manner, by focusing on the corresponding 
aspect of the parent string theory and then studying the effect of the 
limit. An alternative approach would be to take the limit, once and for 
all, at the level of the worldsheet action. This was in fact 
accomplished by Gomis and Ooguri \cite{go}, and has the advantage of 
producing a finite worldsheet Lagrangian which serves as a more explicit 
definition of the full Wound string theory. It is the purpose of this 
paper to derive an analogous membrane worldvolume action for WM2/OM theory.

Our work is in consonance with the idea that it should be possible to 
capture the physics of WM2/OM theory through an appropriate membrane 
action (or some regularized version thereof). This would certainly have 
to be the case if the same statement were true for M theory itself, as 
has been advocated over the years by a number of authors (see 
\cite{nicolai} for some interesting recent developments).
In our more restricted setting, the question is to what extent a 
membrane formulation of WM2/OM theory will be afflicted by the same
problems as the supermembrane. 
We will adopt a pragmatic attitude in this regard, and simply 
try to see how far one can get in developing such a formulation.

The approach of Gomis and Ooguri \cite{go} for deriving the Wound string 
theory action made use of the standard Polyakov action for the string in 
conformal gauge, and involved as an essential ingredient a correlation 
between the worldsheet and longitudinal spacetime light-cones. It is not 
evident how one might generalize their approach
to the membrane case, so we will develop an alternative procedure. For 
simplicity, we will begin by examining the point particle case in 
Section \ref{partsec}, where we will find that the Nambu-Goto type 
description allows for a much simpler derivation of the limiting form of 
the action, and one that can be easily generalized to the 
higher-dimensional cases. We will then apply this approach in Section 
\ref{strsec} to derive a reparametrization-invariant action for Wound 
string theory, which can be gauge-fixed to obtain the known closed and 
open string spectra.

In Section \ref{memsec} we proceed to the membrane case, our main 
interest. We restrict our attention to the bosonic part of the system, 
leaving its supersymmetric completion for future work. We obtain a 
reparametrization-invariant action for WM2 theory,
and examine the description it provides for the various objects of the 
theory. For closed membranes or open membranes ending on a fully 
transverse M5-brane,
a static gauge choice reduces the action to free-field form, and leads 
to a discrete non-relativistic spectrum. Upon double dimensional 
reduction, the closed membrane spectrum is shown to yield the correct 
Wound closed string spectrum.
The cases where the open membrane ends on a partially transverse or 
longitudinal fivebrane are incompatible with the choice of static gauge, 
and therefore qualitatively different (remember that the latter is the 
OM theory setup). We show that in these cases the action can be 
simplified by choosing an orthonormal gauge, and determine the relevant 
boundary conditions. 
The potential which follows from this action is seen to possess flat
directions, implying instabilities.
In addition, we compute the tensions of each of 
the three types of M5-brane in WM2 theory, and find that the fully 
transverse fivebrane is tensionless.
Our conclusions are summarized and discussed in 
Section \ref{conclsec}. We also include an Appendix with a more careful 
analysis, carried out within the Hamiltonian formalism, of the limiting 
and gauge-fixing procedures, which provides a useful complement to the 
perspective of the main text.

\section{The Point Particle}
\label{partsec}

A relativistic point charge of
mass $m$ in a background gauge field $A_{\mu}$ can be described with
the standard action
\be \label{ngpart}
S=-m\int d\tau 
\left[\sqrt{-\dot{X}^{\mu}\dot{X}^{\nu}g_{\mu\nu}}-A_{\mu}\dot{X}^{\mu}\right]
\ee
(where dots stand for $\tau$-derivatives), or equivalently, with
\be \label{ppart}
I=-{m\over 2}\int d\tau 
\left[\sqrt{-\gamma}\left(\gamma^{-1}\dot{X}^{\mu}\dot{X}^{\nu}g_{\mu\nu}+1\right)
-2A_{\mu}\dot{X}^{\mu}\right]~,
\ee
where an instrinsic worldline metric $\gamma$ has been introduced
as an auxiliary variable. The equivalence of (\ref{ngpart}) and 
(\ref{ppart})
can be made manifest by eliminating $\gamma$ from $I$ using its equation 
of motion,
which sets $\gamma$ equal to the induced metric 
$\dot{X}^{\mu}\dot{X}^{\nu}g_{\mu\nu}$.

Let us now study the effect of the $\eps\to 0$ limit
\be \label{partlim}
m=M\eps^{-1}, \qquad g_{\mu\nu}=(-1,\eps,\eps,...), \qquad 
A_0=1-\lambda\eps,
\qquad \mbox{$M$ and $\lambda$ fixed,}
\ee
which is the particle analog of (\ref{strlim}) and (\ref{memlim}).
Consider first the effect on $I$. Inserting (\ref{partlim}) in 
(\ref{ppart}), and choosing for convenience the gauge $\gamma=-1$, we have
\be
I=-{M\over 2\eps}\int d\tau \left[(\dot{X}^0)^2+1-2A_0\dot{X}^0\right]
+{M\over 2}\int d\tau \dot{X}_{\perp}^2~.
\ee
The relative scaling in (\ref{partlim})
  between $m$ and the spatial components of the metric has been chosen 
so as to make the action for the spatial coordinates $X_{\perp}$ 
manifestly finite.
To deal with the divergence seen in the temporal term, we proceed in 
analogy with \cite{go}, rewriting the action by means of a Lagrange 
multiplier $l$,
\be
I=-\int d\tau\left[ l(\dot{X}^0-A_0)-{\eps\over 2M}l^2+1-(A_0)^2
-{M\over 2}\dot{X}_{\perp}^2\right]~.
\ee
The limit $\eps\to 0$ can then be taken without any difficulty, yielding
\be \label{nrppart}
I_{\mbox{\scriptsize W}}=
-\int d\tau\left[l(\dot{X}^0-1)-{M\over 2}\dot{X}_{\perp}^2+\lambda 
M\right]~.
\ee
Notice that, in the end, $l$ is nothing but $p_0$, the momentum 
conjugate to $X^0$ (see the Appendix).
Its equation of motion implies the `static gauge' condition 
$\dot{X}^0=1$, which
is simply our original gauge condition 
$\gamma=\dot{X}^{\mu}\dot{X}^{\nu}g_{\mu\nu}=-1$,
in the limit (\ref{partlim}).
$I_{\mbox{\scriptsize W}}$ is thus seen to be the action for a 
non-relativistic particle of mass $M$. Indeed, demanding that the associated
worldline Hamiltonian vanish (just as it did for the original system,
as a result of reparametrization invariance), 
one finds the energy spectrum $p_0=p_{\perp}^2/2M+\lambda M$.
Its non-relativistic nature is a result of the scaling of the metric in 
(\ref{partlim}), which takes the speed of light to infinity. 
The presence of $A_0$ shifts the momentum conjugate to $X^0$; the  role 
of the gauge field is thus merely to subtract a (dynamically irrelevant) 
divergent contribution to $p_0$, leaving behind a finite energy shift 
controlled by the free parameter $\lambda$.
A more general gauge field configuration $A_0(x)=1-\eps a_0(x)$,
$A_i(x)=\eps a_i(x)$, results in a coupling of the non-relativistic 
particle to the gauge field $a_\mu(x)$.

Now consider the effect of the limit (\ref{partlim}) on 
the action (\ref{ngpart}). With the metric 
scaling according to (\ref{partlim}),
the square root can be expanded in powers of $\eps$. The leading term, 
$\dot{X}^0$, is finite, so multiplied by $m\propto\eps^{-1}$ it would 
imply a divergent contribution to the action. It is precisely cancelled, 
however, by the leading contribution of the gauge field. The subleading 
terms
lead then to the finite action
\be \label{nrngpart}
S_{\mbox{\scriptsize W}}
=-\int d\tau \left[-{M\over 2}{\dot{X}_{\perp}^2 \over 
\dot{X}^0}+\lambda M\dot{X}^0\right]~,
\ee
which can be recognized as the reparametrization-invariant version of 
(\ref{nrppart}), and in particular leads to the same non-relativistic 
spectrum. It is clear then that $S$ is a more convenient starting point 
than $I$ for the purpose of taking the limit (\ref{partlim}): it allows 
for a succinct and transparent derivation of the limiting form of 
the action. Equally important for our purposes is the fact that the 
Nambu-Goto-based approach is readily generalizable to the string and 
membrane cases.
As shown in the Appendix,
the Hamiltonian version of the limit is also completely 
straightforward.

\section{The Wound (Non-relativistic) String}
\label{strsec}

\subsection{The Wound string action}
\label{stractsec}

As noted in the Introduction, the authors of \cite{go} were able 
to derive an action for Wound string theory by starting with the 
Polyakov action for the string, in conformal gauge, and then taking the 
limit (\ref{strlim}). By analogy with the particle case studied in the 
previous section, we expect the analysis of the limit to be more transparent 
if we start instead with the
Nambu-Goto action
\be \label{ngstr}
S=-t_1\int d^2\sigma \left[ \sqrt{-\det 
g_{\mu\nu}\p_{\alpha}X^{\mu}\p_{\beta}X^{\nu}}
-B_{01}\varepsilon^{\alpha\beta}\p_{\alpha}X^{0}\p_{\beta}X^{1}\right]~,
\ee
where $t_1=1/2\pi\ls^2$ is the string tension, $\alpha,\beta=0,1$ are 
worldsheet indices associated with the Lorentzian coordinates 
$\sigma^0\equiv\tau,\sigma^1\equiv\sigma$, and $\varepsilon^{01}=+1$.
As a result of the two-dimensional reparametrization invariance of $S$, 
the worldsheet energy-momentum
tensor (the Noether current for translations) vanishes identically.

For the purpose of taking the limit (\ref{strlim}), it is convenient
to denote the longitudinal and transverse variables by $X^a$ ($a=0,1$) 
and $Y^i$ ($i=2,\ldots,D-1$),
respectively. With the metric scaling according to (\ref{strlim}), the 
longitudinal coordinates give the dominant contribution to the 
Nambu-Goto determinant. This leading term is in fact a perfect square,
\be
-\det \eta_{ab}\p_{\alpha}X^{a}\p_{\beta}X^{b}= 
\left(\varepsilon^{\alpha\beta}\p_{\alpha}X^{0}\p_{\beta}X^{1}\right)^2~.
\ee
Just like in the point-particle case, then,
the leading term in the expansion of the Nambu-Goto square-root in 
(\ref{ngstr}) in powers of \eps\
(which would imply a divergent action, due to $t_1\propto\eps^{-1}$,) is 
precisely cancelled by the leading gauge-field contribution. Notice this 
cancellation takes place only if
$\dot{X}^0 X'^1-\dot{X}^1 X'^0>0$, where primes denote 
$\sigma$-derivatives, i.e., if the string is oriented along $+x^1$: only 
in this case can its diverging tensile energy be compensated by its 
coupling to the background $B$-field.
As long as this condition is satisfied, the subleading terms yield the 
finite action
\be \label{nrngstr}
S_{\mbox{\scriptsize W}}=
-T_1\int d^2\sigma \left[\frac{2\dot{X}\cdot X'\dot{Y}\cdot Y'-\dot{X}^2 
Y'^2-X'^2\dot{Y}^2}
{2\left(\dot{X}^0 X'^1-\dot{X}^1 X'^0\right)}
+\lambda\left(\dot{X}^0 X'^1-\dot{X}^1 X'^0\right)\right]~,
\ee
where $T_1=1/2\pi\Ls^2$ is the effective string tension, and the 
longitudinal and transverse coordinates are respectively contracted with 
  $\eta_{ab}$ and $\delta_{ij}$. This is then the 
reparametrization-invariant worldsheet action of Wound string theory. 
Again, due to this invariance, the worldsheet energy-momentum tensor 
associated with (\ref{nrngstr}) is identically zero.

\subsection{Gauge-Fixing and Spectrum}
\label{strspecsec}

If it is true that the action (\ref{nrngstr}) completely captures the 
dynamics of Wound string theory,
then in particular it should be able to reproduce the spectrum of 
excitations of the various objects in the theory. Besides closed 
strings, which require enforcing the periodicity conditions\footnote{We 
assume for simplicity that only $x^1$ is compact.}
\be \label{cstrbc}
X^a(\sigma+2\pi,\tau)=X^a(\sigma,\tau)+2\pi wR\delta^a_1, \qquad
Y^i(\sigma+2\pi,\tau)=Y^i(\sigma,\tau),
\ee
there may be D-branes in the theory, whose excitations will be 
described as usual by open strings attached to them. 
The action (\ref{nrngstr}) 
implies that boundary conditions must be chosen for these open strings 
such that
\bea \label{strbdry}
\delta X^a\left[\frac{\dot{X}_a\dot{Y}\cdot Y'-X'_a\dot{Y}^2}
{\left(\dot{X}^0 X'^1-\dot{X}^1 X'^0\right)}
-\varepsilon_{ab}\dot{X}^b\left(\frac{2\dot{X}\cdot X'\dot{Y}\cdot 
Y'-\dot{X}^2 Y'^2-X'^2\dot{Y}^2}
{2\left(\dot{X}^0 X'^1-\dot{X}^1 X'^0\right)^2}+\lambda\right) \right]  \\
+\delta Y^i\left[\frac{\dot{Y}_i\dot{X}\cdot X'-Y'_i\dot{X}^2}
{\left(\dot{X}^0 X'^1-\dot{X}^1 X'^0\right)} \right]&=&0~. \nonumber
\eea

For D$p$-branes transverse to $x^1$,
(\ref{strbdry}) implies the standard boundary conditions
\be \label{tstrbc}
X'^0=0, \quad X^1(\sigma=0)=0, \quad X^1(\sigma=\pi)=2\pi wR, \quad 
Y'^j_N=0, \quad \dot{Y}^k_D=0~,
\ee
where $Y_N$ ($Y_D$)
denotes the $p-1$ ($D-p-1$) transverse Neumann (Dirichlet) directions.

The simplicity of (\ref{cstrbc}) and (\ref{tstrbc}) allows for a 
straightforward derivation of the corresponding spectra. The quickest 
route is to note that, since in these cases the string is 
necessarily wound around the longitudinal direction,
we can work in the static gauge $X^0=c\tau$,$X^1=\zeta wR\sigma$,
with $c$ an arbitrary constant and $\zeta=1$ ($\zeta=2$) for the closed 
(open) string. In this gauge, (\ref{nrngstr}) reduces to
the free-field action
\be \label{nrsstr}
S_{\mbox{\scriptsize W}}^{\mbox{\scriptsize (s)}}=
-T_1\int d^2 x \left[{1\over 2}\p_a Y\cdot \p^a Y+\lambda\right]~.
\ee
This is of course
the action for a non-relativistic string.
The usual mode expansions then lead to the expected energy spectra:
\be \label{p0cstr}
p_{0}=\lambda {\frac{wR}{\ensuremath{L_s}^{2}}}+{\frac{\ensuremath{L_s}%
^{2}p_{\perp }^{2}}{2wR}}+{\frac{N+\ensuremath{\tilde{N}}}{wR}}
\ee
for the closed strings \cite{km,wound,go}, and
\be \label{p0tstr}
p_{0}=\lambda{wR\over \Ls^{2}}+{\Ls^{2}p_{\perp}^{2}\over 2wR}
     +{N\over 2wR}
\ee
for the open strings \cite{newt}.

The case of a longitudinal D-brane (i.e., NCOS theory) is qualitatively 
different.
This is to be expected, for it is known to lead to a \emph{relativistic}
open string spectrum \cite{sst2,ncos,wound,go}.
The essential difference is that in this case the endpoint variations
$\delta X^0$ and $\delta X^1$ are both arbitrary, so (\ref{strbdry}) 
implies non-linear
boundary conditions for $X^a$ (namely, the expression inside the first 
pair of
brackets must vanish). These boundary conditions are in fact 
incompatible with the static gauge choice: physically, the point is that 
the length of the string along direction $1$ is not fixed.

To deal with these complicated boundary conditions,
it is convenient to work in conformal gauge.
The resulting formalism will
of course also be able to cover the closed string and transverse D-brane 
cases.
In the standard Polyakov action approach, fixing the conformal gauge means
requiring the \emph{intrinsic}
worldsheet metric to be conformally flat, i.e., 
$\gamma_{01}=0,\gamma_{00}=-\gamma_{11}$.
What is not usually stressed, however, is that, through the equation of 
motion for
$\gamma_{\alpha\beta}$,\footnote{$T_{\alpha\beta}=0$, 
which in the conformal approach 
is enforced as a constraint on the physical states.} this ultimately 
entails that the \emph{induced} metric is also conformally flat:
\be \label{confgauge}
g_{\mu\nu}\dot{X}^{\mu}X'^{\nu}=0, \qquad
g_{\mu\nu}\dot{X}^{\mu}\dot{X}^{\nu}=-g_{\lambda\rho}X'^{\lambda}X'^{\rho}~.
\ee
In the Nambu-Goto treatment of the string, then, fixing the conformal 
gauge means enforcing (\ref{confgauge}). The virtue of these 
conditions is of course
that they linearize the equations of motion and boundary conditions
following from (\ref{ngstr}),
while maintaining spacetime covariance.
In addition, the gauge transformations that conditions (\ref{confgauge}) 
leave unfixed,
i.e., conformal reparametrizations, play a central role in the 
development of the formalism (e.g.,
they allow the calculation of scattering amplitudes in terms of vertex 
operators).

Upon taking the limit (\ref{strlim}), the conformal gauge conditions 
(\ref{confgauge}) involve only the longitudinal coordinates $X^a$:
\be \label{nrconfgauge}
\dot{X}\cdot X'=0, \quad  \dot{X}^2=-X'^2
\qquad \Longrightarrow \qquad
X'^0=\dot{X}^1, \quad X'^1=\dot{X}^0.
\ee
In this gauge, the equations of motion following from (\ref{nrngstr}) 
become simple wave equations, and the boundary conditions implicit in 
(\ref{strbdry}) reduce to
\be \label{lstrbc}
\dot{Y}^2+Y'^2=\lambda\left(-\dot{X}^2+X'^2\right)~.
\ee
At this point one must remember the fact that, for longitudinal 
D-branes, the possible presence of a worldvolume electric field 
$F_{01}$, together with the flux quantization condition for 
$B_{01}+2\pi\ls^2 F_{01}$, effectively fixes $\lambda= 1/2\nu^2\Gs^2$, 
where
\be \label{nu}
\nu\equiv {N\Ls^{p-1}\over R_2 \cdot\cdot\cdot R_p}
\ee
is the number density of fundamental strings bound to the D$p$-brane, 
and \Gs\ is the Wound string coupling \cite{newt}.
This is in sharp contrast with the closed string and transverse D-brane 
cases, where $\lambda$ is a strictly free (and physically irrelevant) 
parameter which may for instance be set equal to zero.

Using (\ref{nrconfgauge}), the Wound string theory action 
(\ref{nrngstr}) can be rewritten as
\bea \label{nrpstr}
S_{\mbox{\scriptsize W}}^{\mbox{\scriptsize (c)}}&=&
-T_1\int d^2\sigma \left[-{1\over 2}\dot{Y}^2+{1\over 2} Y'^2
+\lambda\left(\dot{X}^0 X'^1-\dot{X}^1 X'^0\right) \right. \nonumber\\
{}&{}& \left. \qquad\qquad\quad
+l_0\left(\dot{X}^0-X'^1\right)+l_1\left(\dot{X}^1-X'^0\right) \right]~,
\eea
where Lagrange multipliers $l_a$ have been introduced to
appropriately enforce the gauge conditions (\ref{nrconfgauge}).
A more careful Hamiltonian justification of this gauge-fixing procedure 
is given in the Appendix.
The action (\ref{nrpstr}) yields the equations of motion
\be \label{nrstreom}
\ddot{Y}=Y'', \quad X'^0=\dot{X}^1, \quad X'^1=\dot{X}^0,
\quad l'_0=\dot{l}_1, \quad l'_1=\dot{l}_0
\ee
(so, in particular, $X^a, l_a$ satisfy the wave equation), and requires 
boundary conditions such that
\be \label{confstrbdry}
\delta Y^i Y'_i+\delta 
X^a\varepsilon_{ab}\left[\lambda\dot{X}^b+l^b\right]=0.
\ee
For the longitudinal D-brane (i.e., NCOS) case, this translates into
the linear boundary conditions
\be \label{conflstrbc}
Y'^j_N=0, \quad \dot{Y}^k_D=0, \quad l_0=\lambda\dot{X}^0, \quad 
l_1=-\lambda\dot{X}^1.
\ee
For points on the boundary, the action (\ref{nrpstr}) implies the 
two-point function
\be \label{nc2pt}
\expec{X^a(\tau)X^b(0)}=-\nu^2\Gs^2\Ls^2\eta^{ab}\log|\tau|^2+
   \pi\nu^2\Gs^2\Ls^2\varepsilon^{ab}\mbox{sgn}(\tau)~,
\ee
which exhibits the expected noncommutativity between space and time,
with noncommutativity parameter
$\theta^{ab}=2\pi\nu^2\Gs^2\Ls^2\varepsilon^{ab}$ \cite{sw,sst2,ncos,go}.
Other ways to make this noncommutativity apparent
have been explored in \cite{ncstr}.\footnote{In this connection, we would
like to stress that regarding boundary conditions as Hamiltonian constraints is a procedure whose consistency remains to be established.}

In order for (\ref{nrpstr}) to be equivalent to (\ref{nrngstr}), we must 
require the worldsheet energy-momentum tensor $T_{\alpha\beta}$
which follows from $S_{\mbox{\scriptsize W}}^{\mbox{\scriptsize (c)}}$ 
to vanish, just as it did for the original system. This translates into 
the two independent constraints
\bea \label{nrconfconst}
\dot{Y}\cdot Y'-l_a X'^a &=& 0, \\
\dot{Y}^2+ Y'^2-2\left(l_0 X'^1+l_1 X'^0\right) &=& 0 \nonumber
\eea
Equivalence to the original system (in conformal gauge) can be shown 
explicitly by solving the constraints (\ref{nrconfconst}) for the 
Lagrange multipliers. After a slight rewriting using (\ref{nrstreom}), 
this leads to
\be \label{lasol}
l_a=\frac{\dot{Y}\cdot Y' X'_a - {1\over 
2}\left(\dot{Y}^2+Y'^2\right)\dot{X}_a}
{\dot{X}^0 X'^1-\dot{X}^1 X'^0}.
\ee
Direct substitution of these solutions into (\ref{conflstrbc}) takes us 
back to the non-linear boundary condition (\ref{lstrbc}). It is much 
simpler, however, to retain the auxiliary variables $l_a$ in the 
description, expand all variables into modes which satisfy 
(\ref{nrstreom}) and (\ref{conflstrbc}), and enforce the constraints 
(\ref{nrconfconst}) only a posteriori, as physical state conditions. The 
zero-mode of the Hamiltonian constraint can then easily be seen to yield 
the expected NCOS spectrum \cite{newt}
\be \label{p0long}
\nu^{2}\Gs^{2}
   \left[ (p_{0})^{2}-(p_{1})^{2}\right]- p_{\perp}^{2}=
   {N\over \Ls^{2}}~.
\ee

That (\ref{nrpstr}) leads to the correct spectrum for open strings 
attached to longitudinal branes is not a new result: letting 
$2\beta\equiv-l_0-l_1$, $2\tbeta\equiv-l_0+l_1$, $\gamma\equiv X^0+X^1$, 
and
$\tgamma\equiv-X^0+X^1$, $S_{\mbox{\scriptsize W}}^{\mbox{\scriptsize 
(c)}}$ is seen to be just a (Lorentzian) rewriting of the Gomis-Ooguri 
action \cite{go}, which has been previously shown to yield the right 
spectra for closed strings \cite{go} as well as longitudinal 
\cite{go,newt} and transverse D-branes \cite{newt}.

Although in (\ref{nrpstr}) we have arrived at a known result,
we believe our approach sheds some additional light
on the worldsheet-level effect of the Wound string theory limit.
Most importantly, as in the case of the point particle, we have seen in 
Section \ref{stractsec}
that the use of the Nambu-Goto action (\ref{ngstr}) as a starting point 
makes the cancellation between tensile and Kalb-Ramond potential 
energies which is the physical essence of the Wound string theory limit
completely transparent.
At a more technical level, the analysis of this subsection clarifies a 
couple of aspects of the Gomis-Ooguri action: besides showing that the 
respectively purely left- and purely right-moving character of 
$\gamma=X^+$ and $\tgamma=X^-$ \cite{sst2,go} is not a dynamical effect, 
but a direct consequence of the conformal gauge conditions 
(\ref{nrconfgauge}), it gives a physical meaning (through 
Eq.~(\ref{lasol})) to the Lagrange multipliers $\beta,\tbeta$ appearing 
in the formalism of \cite{go}.

\newpage

\section{The Wrapped (Galilean) Membrane}
\label{memsec}

\subsection{The WM2 action}
\label{memactsec}

The bosonic part of the action for an M2-brane in a background $C_{012}$ 
field can be written in the Nambu-Goto form\footnote{Supermembrane 
reviews may be found in \cite{supermembrane,matrix}.}
\be \label{ngmem}
S=-t_2\int d^3\sigma \left[ \sqrt{-\det 
g_{MN}\p_{\alpha}X^{M}\p_{\beta}X^{N}}
-C_{012}\varepsilon^{\alpha\beta\gamma}\p_{\alpha}X^{0}\p_{\beta}X^{1}\p_{\gamma}X^{2}\right]~,
\ee
where $t_2=1/(2\pi)^2\lP^3$ is the membrane tension, the worldvolume 
coordinates $\sigma^{\alpha}\equiv(\tau,\sigma,\rho)$, 
$\varepsilon^{012}=+1$, and the spacetime indices $M,N=0,\ldots,10$.
With this action as our starting point, the limit (\ref{memlim}) can be taken in 
complete parallel with the particle and string cases analyzed in Section 
\ref{strsec}.\footnote{It is also straightforward to generalize to 
arbitrary $p$-branes.}
It is convenient again to make a notational distinction between 
longitudinal and transverse coordinates: $X^a$ ($a=0,1,2$), $Y^i$ 
($i=3,\ldots,10$). The key point is once more that the Nambu-Goto 
square-root is dominated by the longitudinal piece of the determinant,
\be
\sqrt{-\det \eta_{ab}\p_{\alpha}X^{a}\p_{\beta}X^{b}}=
|\varepsilon^{\alpha\beta\gamma}\p_{\alpha}X^{0}\p_{\beta}X^{1}\p_{\gamma}X^{2}|~.
\ee
If the expression within the absolute value is positive (i.e., if 
the membrane is positively oriented on the $x^1$-$x^2$ torus), these 
terms cancel
against the leading gauge-field terms, and we are left with the finite 
action
\bea \label{nrngmem}
S_{\mbox{\scriptsize W}}&=&-T_2\int d^3\sigma 
\left[-\frac{\varepsilon^{\alpha\beta\gamma}\varepsilon^{\alpha'\beta'\gamma'}
\p_{\alpha}X\cdot\p_{\alpha'}X
\p_{\beta}X\cdot\p_{\beta'}X
\p_{\gamma}Y\cdot\p_{\gamma'}Y}
{4\varepsilon^{\alpha\beta\gamma}
\p_{\alpha}X^{0}\p_{\beta}X^{1}\p_{\gamma}X^{2}} \right. \nonumber\\
{}&{}& \qquad\quad\qquad\left.+\lambda
\varepsilon^{\alpha\beta\gamma}
\p_{\alpha}X^{0}\p_{\beta}X^{1}\p_{\gamma}X^{2}
\right]~,
\eea
where $T_2=1/(2\pi)^2\LP^3$ is the effective membrane tension,
and the longitudinal and transverse coordinates are respectively 
contracted with  $\eta_{ab}$ and $\delta_{ij}$. Eq.~(\ref{nrngmem}) is 
the reparametrization-invariant worldvolume
action of Wrapped membrane theory. As we will explain later, for the OM 
theory case
(where the membrane is open and ends on a longitudinal fivebrane) the 
non-linear self-duality constraint on the M5 worldvolume leads to an 
additional term in the action, Eq. (\ref{omextra}).
The invariance of $S_{\mbox{\scriptsize W}}$ under diffeomorphisms 
implies an identically vanishing
worldvolume energy-momentum tensor.

Just like in the analysis of the string, there are several cases to 
consider, depending on whether the membrane is closed or open, and if 
open, whether it ends on a longitudinal, partially transverse, or fully 
transverse M5-brane. In the following subsections we will analyze these 
cases separately.

\subsection{Closed membrane spectrum}
\label{cmemsec}

For the closed membrane (parametrized such that $\sigma,\rho$ range from 
$0$ to $2\pi$),
the simplest approach is again to note that, since 
in this case
the membrane is necessarily wrapped around the longitudinal directions, we can
work in the static gauge
\be \label{sgauge}
X^0=c\tau, \quad X^1=\zeta w_1 R_1\sigma, \quad X^2=\zeta w_2 R_2\rho,
\ee
with $c$ an arbitrary constant and $\zeta=1$,
thereby reducing (\ref{nrngmem})
to the free-field action
\be \label{nrsmem}
S_{\mbox{\scriptsize W}}^{\mbox{\scriptsize (s)}}=
-T_2\int d^3 x \left[{1\over 2}\p_a Y\cdot \p^a Y+\lambda\right]~,
\ee
which describes a non-relativistic membrane.
The orientation requirement on the membrane (necessary to arrive at the 
finite action (\ref{nrngmem})) translates into the condition $w\equiv 
w_1 w_2 >0$: the membrane must have positive wrapping number $w$ on the 
longitudinal torus \cite{wound}.\footnote{It should be remembered that, 
in contrast with the individual winding numbers $w_1,w_2$, the wrapping 
number $w$ is conserved by the closed membrane interactions.}
It is straightforward to expand in Fourier modes:
\be \label{smemmodes}
Y^i(x^a)=y^i+{\LP^3\over w R_1 R_2}p_i x^0
+\sqrt{{\LP^3\over w R_1 R_2}}{\sum_{\vec{n}\neq 0}}
\left({a^i_{\vec{n}}\over\sqrt{2q_0}} e^{-iq_a x^a}
+{a^{i\dagger}_{\vec{n}}\over\sqrt{2q_0}} e^{iq_a x^a} \right)~,
\ee
where $\vec{n}\equiv(n_1,n_2)$, $q_a\equiv n_a/w_a R_a$ for $a=1,2$, and
$q_0\equiv\sqrt{q_1^2+q_2^2}$. The resulting energy spectrum is
\be \label{p0cmem}
p_{0}=\lambda {wR_1 R_2\over\LP^3}
+{\LP^3 p_{\perp}^2\over w R_1 R_2}
+{\cN\over wR_1 R_2}~,
\ee
where we have defined a number operator
\be \label{calN}
\cN\equiv\sum_{\vec{n}}\sqrt{(n_1 w_1 R_1)^2+(n_2 w_2 R_2)^2}
a^{\dagger}_{\vec{n}}\cdot a_{\vec{n}}~,
\ee
omitting a possible ordering constant.

It is well-known that a relativistic membrane leads to a continuous 
spectrum \cite{wln}.
This can be understood intuitively from the fact that the Nambu-Goto 
action (\ref{ngmem})
implies an energy proportional to the membrane area, and therefore allows
the membrane to develop arbitrarily long spikes of infinitesimal area,
at zero energy cost.\footnote{There is some disagreement in the 
literature as to whether or not this instability is eliminated upon 
wrapping the membrane on a torus
\cite{russo,rt,dwpp}.}
The formulation of the
Matrix model of M-theory \cite{bfss},
originally obtained as a discretization of the supermembrane action 
\cite{whn},
turned this membrane instability into a virtue:
it is a sign
that the quantized membrane yields a second-quantized description,
with a spectrum that includes multiple-particle
states.\footnote{For reviews on the supermembrane-Matrix connection see, 
e.g., \cite{matrix}.}
An $n$-particle state is obtained by deforming the membrane into 
$n$ blobs connected by infinitesimally thin tubes, which carry no energy.
In this way, a single membrane leads to configurations which are indistinguishable
from multiple-membrane states. 
In contrast with this standard case, we have found here that the Wrapped 
membrane action
(\ref{nrngmem}) implies a \emph{discrete} spectrum. This is of course 
due to its non-relativistic character: as is evident in (\ref{nrsmem}), 
the membrane potential does not have flat directions.
What has happened, then, is that, as a consequence of the limit
(and, as we will see later, our choice of gauge), the
connecting tubes mentioned above carry a high energy cost, and are therefore forbidden,
implying that the multi-particle states are removed from the spectrum.
As stated before,
all states in the resulting spectrum carry strictly positive M2-brane
wrapping number $w$, which in particular means that 
the limit (\ref{memlim}) decouples the massless supergravity modes 
\cite{om,bbss2} (although these remain in the theory as carriers of 
Newtonian interactions, as described in \cite{newt}).

As a check on the closed membrane spectrum (\ref{p0cmem}), notice that 
under (double) dimensional reduction along $x_2$ (i.e., letting $n_2=0$, 
$w_2=1$) it agrees with the closed string spectrum (\ref{p0cstr}), with 
the expected identifications
\be \label{memtostr}
\Ls^2={\LP^3\over R_2}, \quad
N=\sum_{{n_1>0}\atop{n_2=0}} n_1 a^{\dagger}_{\vec{n}}\cdot a_{\vec{n}}, 
\quad
\tN=\sum_{{n_1<0}\atop{n_2=0}} |n_1| a^{\dagger}_{\vec{n}}a_{\vec{n}}~.
\ee

\subsection{Fivebranes and open membranes}
\label{omemsec}

It is well-known that open M2-branes can terminate consistently on 
M5-branes \cite{strominger,townsend}.
By analogy with the case of open strings ending on D-branes, it is 
natural to expect the fivebrane dynamics to be describable through an 
appropriate open membrane quantization \cite{bb,bm,emm2,dwpp2,cs}.
In the present subsection we will address this issue in the more 
restricted WM2 theory context. Before doing that, however, it is worth 
establishing some basic M5-brane properties which follow directly from 
the nature of the WM2 limit.

The scaling of the metric seen in (\ref{memlim}) introduces a 
distinction between the $x^1,x^2$ (longitudinal) and $x^3,\ldots,x^{10}$ 
(transverse) directions, so the M5-brane attributes become 
embedding-dependent. There are three qualitatively distinct cases to 
consider: the fivebrane may be fully transverse (e.g., spanning 
directions $034567$), partially transverse (e.g., $023456$), or 
longitudinal (e.g., $012345$).
In the first case, the usual formula for the M5 tension implies in the 
limit (\ref{memlim}) an energy per unit coordinate volume (i.e., proper 
volume in the WM2 metric)
\be \label{tt5}
T^{\perp\perp}_{M5}=\frac{(g_{\perp\perp})^{5/2}}{(2\pi)^5\lP^6}
    =\frac{\sqrt{\eps}}{(2\pi)^5\LP^6}\to 0~.
\ee
That is to say, fully transverse M5-branes in WM2 theory are 
tensionless! This is consistent with the fact that, upon reduction along 
$x^2$ (say), a fully transverse M5-brane becomes a transverse NS5-brane 
in Wound IIA (WIIA) string theory, an object which, as mentioned in the 
Introduction, is also known to be tensionless.\footnote{Through duality 
one can similarly infer that transverse Kaluza-Klein fivebranes in DLCQ 
string theory, and transverse D$(6-p)$-branes in Wrapped D$p$-brane 
theory (which is related to the $(p+3)$-dimensional NCYM and 
$6$-dimensional OD$p$ theories \cite{wound}), among others, are also 
tensionless \cite{talk}.}

For a partially transverse M5-brane, the same reasoning implies a finite 
tension
\be \label{lt5}
T^{\para\perp}_{M5}=\frac{(g_{\perp\perp})^{4/2}}{(2\pi)^5\lP^6}
    =\frac{1}{(2\pi)^5\LP^6}~.
\ee
Reducing along $x^2$ ($x^1$), an M5-brane oriented along $023456$ 
becomes a transverse D4-brane (longitudinal NS5-brane) in WIIA string 
theory. Remembering that the WM2 and WIIA parameters are related through
\be \label{wm2wiia}
\LP=\Gs^{1/3}\Ls, \qquad R_{2}=\Gs\Ls \quad (R_{1}=\Gs\Ls),
\ee
one can verify that indeed
\be \label{t4}
T^{\para\perp}_{M5}2\pi R_2=\frac{1}{(2\pi)^4\Gs\Ls^5}
    =T^{\perp}_{D4}~,
\ee
the tension of a transverse D4-brane \cite{wound}, and
\be \label{lNS5}
T^{\para\perp}_{M5}=\frac{1}{(2\pi)^5\Gs^2\Ls^6}
    =T^{\para}_{NS5}~,
\ee
the tension of a longitudinal NS5-brane \cite{br,talk}.

In the longitudinal case, the formula directly analogous to 
(\ref{tt5}),(\ref{lt5}) would yield a divergent fivebrane tension:
$(g_{\perp\perp})^{3/2}/(2\pi)^5\lP^6\propto 1/\sqrt{\eps}$. This just 
says that an \emph{isolated} longitudinal M5-brane does not survive the 
WM2 limit. To remain in the spectrum of the theory, the fivebrane must 
be bound to some number $w>0$ of longitudinal M2-branes. In other words, 
it must carry a positive $\cF_{012}$ field, where $\cF_{mnp}\equiv 
C_{mnp}+H_{mnp}$ (with $C_{mnp}$ the pull-back of the bulk gauge field, 
$H_{mnp}$ the worldvolume field-strength, and $m,n,p=0,\ldots,5$) is the 
gauge invariant three-form field. $\cF_{012}$ and $w$ are related 
through a flux-quantization condition. For fixed $w$, and with the 
metric and Planck length scaling as in (\ref{memlim}), this condition 
implies that $\cF_{012}$ must become near-critical, as seen for 
$C_{012}$ in (\ref{memlim})
(we work in the gauge $H_{mnp}=0$).\footnote{The entire analysis here is 
a direct analog of the D$p$-F1 (NCOS) case, discussed in more detail in 
Sec. 3.2.1 of \cite{newt}.} In short, longitudinal M5-branes in WM2 
theory give rise to the standard OM theory setup \cite{wound}.
For later use, it is important to remember that the non-linear 
self-duality constraint \cite{selfdual}
for $\cF$ implies that the `electric' component $C_{012}$ seen in 
(\ref{memlim}) must be accompanied by a `magnetic' counterpart 
\cite{om,bbss2}
\be \label{c345}
C_{345}=-{\eps\over\sqrt{2\lambda}}~.
\ee

The bottom line of the preceding discussion is that, for longitudinal 
fivebranes, one must examine the behavior of the M5-M2 bound state 
tension in the limit (\ref{memlim}):
\be \label{m5m2}
\sqrt{\left[\frac{1}{(2\pi)^2\lP^3}\frac{w}{(2\pi)^3 R_3 R_4 R_5}\right]^2
+\left[\frac{(g_{\perp\perp})^{3/2}}{(2\pi)^5\lP^6}\right]^2}
\quad\longrightarrow\quad
\frac{1}{(2\pi)^5\LP^6}\left({|\upsilon|\over\eps}+{1\over 
2|\upsilon|}\right)~,
\ee
where
\be \label{upsilon}
\upsilon\equiv {w\LP^3\over R_3 R_4 R_5}
\ee
is the number of M2-branes (oriented along 012) per unit transverse 
volume on the M5-brane (oriented along 012345). The leading term in 
(\ref{m5m2}) is divergent, but the divergence is proportional to the 
membrane wrapping number $w$ carried by the fivebrane. Since the total 
wrapping number (including both free M2 and bound M2 contributions) is 
conserved by interactions, the above $\eps^{-1}$ divergence will be 
dynamically irrelevant as long as all objects carry strictly positive 
$w$ \cite{wound}. In fact, with our choice of gauge for $\cF$ (i.e., 
when it is $C$ and not $H$ that becomes near-critical), this divergence 
is automatically subtracted by the coupling to the $C_{012}$ field 
(\ref{memlim}). {}From the subleading term in (\ref{m5m2}) we conclude 
then that the tension of a longitudinal M5-brane in WM2 theory is given by
\be \label{ll5}
T^{\para\,\para}_{M5}=\frac{1}{2(2\pi)^5\upsilon\LP^6}~.
\ee
If we reduce to ten dimensions along $x^2$ (say), then using (\ref{nu}), 
(\ref{wm2wiia}) and (\ref{upsilon}) we can readily verify that
\be \label{l4}
T^{\para\,\para}_{M5}2\pi 
R_2=\frac{1}{2(2\pi)^4\nu\Gs^2\Ls^5}=T^{\para}_{D4}~,
\ee
which is the correct tension for a longitudinal D4-brane in Wound IIA 
string theory \cite{wound}.

Let us now proceed to determine what information may be extracted from 
the WM2 action (\ref{nrngmem}) when the membrane is open and ends on one 
of the above types of M5-brane. We should note that, as has been 
emphasized in \cite{bbss1}, the absence of a dimensionless coupling 
constant in M-theory implies that in general there is no sense in which 
the M2-brane tension may be regarded as small compared to the M5-brane 
tension (which one can do for F1 vs. D$p$ in string theory, as long as 
$\gs\ll 1$). This means that, \emph{a priori}, it is far from clear 
whether the M5-brane may be considered a rigid wall on which the open 
M2-brane terminates.
Still, if one adopts the view that the physics of M theory is completely 
captured by membrane quantization--- as we do here for WM2 theory--- 
then the question is simply what boundary conditions are consistently 
allowed by the relevant action.\footnote{Notice also that, in contrast 
to M theory, WM2 theory does include the dimensionless parameters 
$R_{1}/\LP$, $R_{2}/\LP$ and $\upsilon$, which enter into the comparison 
of the M2 and M5 tensions.}
This is the issue which we now address.

We take $\sigma,\rho$ to range from $0$ to $\pi$.
For an open membrane ending on a fully transverse fivebrane, variation 
of (\ref{nrngmem}) leads to a boundary term which can be seen to allow 
the `obvious' boundary conditions
\bea \label{ttbc}
{}&{}&\sigma=0,\pi:\qquad X'^0=0, \quad X^{1,2}=c^{1,2}, \quad Y'^j_N=0, 
\quad Y^k_D=d^k, \\
{}&{}&\rho=0,\pi:\qquad \hat{X}^0=0, \quad X^{1,2}=c^{1,2}, \quad 
\hat{Y}^j_N=0, \quad Y^k_D=d^k, \nonumber
\eea
where primes and hats denote $\sigma$- and $\rho$-derivatives, 
respectively, and $c^a,d^k$ are constants.
Our bosonic analysis does not constrain the number of transverse Neumann 
($Y^j_N$) and Dirichlet ($Y^k_D$) directions.
For a fivebrane in ten spatial dimensions we would normally expect five 
of the $Y^i$ to be Neumann, and three of them (together with $X^{1,2}$) 
to be Dirichlet. We have seen above, however, that the M5-brane in 
question is tensionless, which means that it can be deformed to any 
shape in the transverse space, at zero energy cost. We should perhaps 
anticipate then that this object will be effectively eight-dimensional, 
and all of the $Y^i$ will satisfy Neumann boundary 
conditions.\footnote{A similar situation arises in the analysis of open 
string field theories for unstable D-brane systems: at the closed string 
vacuum the branes are tensionless, and so effectively become 
space-filling, which explains why closed strings (understood to be flux 
tubes of the D-brane worldvolume gauge field) can move about freely in 
the nine-dimensional bulk \cite{fluxstrings}.}
Settling this issue will require an analysis of the supersymmetric 
completion of the WM2 action, and in particular the compatibility 
between boundary conditions and $\kappa$-symmetry, a question which has 
been studied in the full M theory setting in 
\cite{bm,emm2,dwpp2}.\footnote{It was found in that context that the 
M2-brane may terminate on a $p$-brane only if $p=1$(!), 5, or 9.} We 
leave this interesting problem for future work.

Irrespective of that, we can progress further by noting that the 
boundary conditions (\ref{ttbc}) are compatible with the static gauge 
(\ref{sgauge}), with $\zeta=2$. The open membrane action can therefore 
be cast again in the free-field form (\ref{nrsmem}), implying a discrete 
non-relativistic spectrum essentially identical to (\ref{p0cmem}).
Reduction to ten dimensions yields a prediction for the excitation 
spectrum of a transverse NS5-brane in WIIA theory (since this brane is 
also tensionless, it could again be expected to be effectively 
eight-dimensional).

Unsurprisingly, the static gauge turns out to be inadequate for dealing 
with the partially transverse and longitudinal fivebrane cases. Just 
like in the corresponding string case, the static gauge conditions are 
incompatible with the relevant boundary conditions. Physically, the 
issue is that, in these two cases, the length of the membrane along 
$x^1$ and/or $x^2$ is not fixed.
These setups are thus qualitatively different from the closed membrane 
and transverse fivebrane, which are manifestly non-relativistic. Given 
the relation between OM and $4+1$ NCOS theory, we expect the system to 
have a relativistic structure in the longitudinal fivebrane case. As 
noted before, the partially transverse M5-brane gives rise either to a 
transverse D4-brane (known to possess a non-relativistic spectrum) or a 
longitudinal NS5-brane in WIIA string theory.

A convenient gauge choice which is able to cover all cases is the 
membrane analog of the conformal gauge (\ref{confgauge}) for the string,
\bea \label{orthogauge}
{}&{}& g_{MN}\dot{X}^M X'^N=0,\qquad  g_{MN}\dot{X}^M \hat{X}^N=0,\\
{}&{}& L^2 
g_{MN}\dot{X}^M\dot{X}^N=g_{MN}X'^{M}\hat{X}^{N}g_{PQ}X'^{P}\hat{X}^{Q}
-g_{MN}X'^{M}X'^{N}g_{PQ}\hat{X}^{P}\hat{X}^{Q}, \nonumber
\eea
where $L$ is an arbitrary constant with dimension of length.
We will refer to conditions (\ref{orthogauge}) as orthonormal gauge;
they are advantageous because they eliminate the square-root in 
(\ref{ngmem}).
In the limit (\ref{memlim}), the orthonormal gauge conditions reduce to
\be \label{nrorthogauge}
\dot{X}\cdot X'=\dot{X}\cdot\hat{X}=0,\quad
L^2\dot{X}^2=(X'\cdot\hat{X})^2-X'^2\hat{X}^2
\qquad \Longrightarrow \qquad
L\dot{X}^a=\varepsilon^{abc}X'_b\hat{X}_c~.
\ee
In this gauge, the WM2 action (\ref{nrngmem}) simplifies to
\bea \label{nromem}
S_{\mbox{\scriptsize W}}^{(o)}&=&
-T_2 \int d^3\sigma \left[-\half L\dot{Y}^2
+\half L^{-1}(X'^2\hat{Y}^2-2X'\cdot\hat{X}Y'\cdot\hat{Y}+\hat{X}^2 
Y'^2) \right. \\
{}&{}&\qquad\qquad\quad +\left. 
l_a(L\dot{X}^a-\varepsilon^{abc}X'_b\hat{X}_c)
+\lambda\varepsilon^{\alpha\beta\gamma}
\p_{\alpha}X^0\p_{\beta}X^1\p_{\gamma}X^2
\right]~, \nonumber
\eea
where $l_a$ are Lagrange multipliers enforcing the gauge conditions
(the Appendix shows how this result can be justified using the 
Hamiltonian formalism).
To ensure equivalence with (\ref{nrngmem}), we must demand that the
energy-momentum tensor $T_{\alpha\beta}$ following from (\ref{nromem}) 
vanish. This leads to the three independent constraints
\bea \label{nrorthoconst}
\dot{Y}\cdot Y' - l_a X'^a&=&0, \\
\dot{Y}\cdot\hat{Y}- l_a\hat{X}^a&=&0, \nonumber\\
L^2\dot{Y}^2+Y'^2\hat{Y}^2-(Y'\cdot\hat{Y})^2-2L\varepsilon_{abc}l^a 
X'^b\hat{X}^c&=&0. \nonumber
\eea

For the OM theory (i.e., longitudinal fivebrane) case, the presence of 
the `magnetic' field (\ref{c345}) implies an additional contribution
\be \label{omextra}
-T_2\int d^3\sigma\, 
{1\over\sqrt{2\lambda}}\varepsilon^{\alpha\beta\gamma}\p_{\alpha}Y^3\p_{\beta}Y^4\p_{\gamma}Y^5
\ee
  to the action (\ref{nromem}).
Moreover, in this case the flux quantization condition on the fivebrane 
effectively determines $\lambda$ in terms of the number $w$ of M2-branes 
in the M2-M5 bound state that the flux-carrying M5-brane represents. 
Indeed, by reducing to ten-dimensions along $X^2$, one can use the 
quantization condition for $B_{01}\equiv C_{012}$ (see, e.g., 
\cite{om,newt}) to show that $\lambda=1/2\upsilon^2$, where $\upsilon$ 
is the M2 brane number density defined in (\ref{upsilon}). This is 
completely unlike the closed membrane and (fully or partially) 
transverse fivebrane cases, where $\lambda$ is a free (and dynamically 
inconsequential) parameter.

As a check on (\ref{nromem}), notice that under the double dimensional 
reduction $X^2=\zeta \rho R_2$,
$\hat{X}^{0,1}=\hat{Y}^i=0$, and making use of (\ref{wm2wiia}), 
$S_{\mbox{\scriptsize W}}^{(o)}$
  coincides with the Wound string theory action (\ref{nrpstr}), as long 
as we set $L=\zeta R_2$.\footnote{This is recognized to be the condition 
under which the orthonormal and conformal gauges (\ref{orthogauge}) and 
(\ref{confgauge}) agree.}
As usual, this formal reduction
does not necessarily imply that the membrane will dynamically reduce to a string when the 
circle is shrunk to zero size \cite{russo,rt,dwpp}. This is closely related
to the question of whether or not the membrane displays instabilities.

As mentioned before, the potential for the
standard supermembrane allows arbitrarily long spikes to
develop at zero energy cost. To examine this issue here, consider a static
configuration where, up to small corrections,
all of the spatial variables depend on the same linear combination of 
$\sigma$ and $\rho$ (i.e.,
$X'^r=c\hat{X}^r+\delta$ for $r=1,2$; $Y'^i=c\hat{Y}^i+\delta$), describing a spike
of infinitesimal width. It is then
easy to see that the potential 
energy implied by either (\ref{nromem}) or (\ref{nrngmem}) is of order $\delta$, 
meaning that there are flat directions and instabilities.
At the quantum level, and for the fully supersymmetric system,
we would expect these classical instabilities to translate into a 
continuous spectrum, indicative of multi-particle states.
This seems to be in conflict with our derivation of 
a discrete spectrum, Eq.~(\ref{p0cmem}),
for the closed membrane. The origin of the discrepancy is
the fact that the static gauge (\ref{sgauge}) employed there is incompatible
with the condition $X'^r=c\hat{X}^r+\delta$,
 which as we saw above defines the flat directions
of the potential. We thus see that this gauge choice removes from
the theory the modes responsible
for the instability, and consequently,
 the possiblity to have multi-particle states.

A variation of (\ref{nromem}) $+$ (\ref{omextra}) produces a boundary 
term which on the $\sigma=0,\pi$ edges reads
\bea \label{omembdry}
\delta X^a \left[L^{-1}(X'_a \hat{Y}^2-\hat{X}_a Y'\cdot\hat{Y})
+\varepsilon_{abc}l^b\hat{X}^c+\lambda\varepsilon_{abc}\dot{X}^b\hat{X}^c 
\right]&{}&{} \\
+\delta Y^d \left[L^{-1}(Y'_d \hat{X}^2-\hat{Y}_d X'\cdot\hat{X})
   -\left\{ 
{1\over\sqrt{2\lambda}}\varepsilon_{def}\dot{Y}^e\hat{Y}^f\right\}\right]&{}&{}
\nonumber \\
+\delta Y^l \left[L^{-1}(Y'_l \hat{X}^2-\hat{Y}_l 
X'\cdot\hat{X})\right]&=&0~, \nonumber
\eea
where $a,b,c=0,1,2$; $d,e,f=3,4,5$; $l=6,\ldots,10$. The term inside the 
curly braces comes from (\ref{omextra}), and is
therefore only present in the OM theory case.

For the case of a fully transverse M5-brane, we expect $X^{1,2}=c^{1,2}$ 
at $\sigma=0,\pi$, which
of course implies that $\delta X^{1,2}=0$ and 
$\dot{X}^{1,2}=\hat{X}^{1,2}=0$ there. We then see that
(\ref{omembdry}) (together with the analogous term at the $\rho$-edges)
implies the boundary conditions (\ref{ttbc}), as was claimed previously.
We have already determined the excitation spectrum for this case using 
the static gauge (\ref{sgauge})
(which is of course a subgauge of the orthonormal gauge 
(\ref{orthogauge})), so we will not pursue it further here, except to 
note that, in an orthonormal gauge treatment, the boundary conditions 
for the auxiliary fields $l_a$ would be read off from the constraints 
(\ref{nrorthoconst}).

For a partially transverse M5-brane we expect $X^1$ (but not $X^2$) to 
be Dirichlet. The boundary term (\ref{omembdry}) can then be seen to 
imply the conditions
\bea \label{ltbc}
\sigma=0,\pi:\quad X'^0 \hat{Y}^2=L l^1\hat{X}^2, \quad X^{1}=c^{1}, \quad
X'^2\hat{Y}^2=L l^1\hat{X}^0, \quad Y'^j_N=0, \quad Y^k_D=d^k, \\
\rho=0,\pi:\quad \hat{X}^0 Y'^2=L l^1 X'^2, \quad X^{1}=c^{1}, \quad
\hat{X}^2 Y'^2=L l^1 X'^0, \quad \hat{Y}^j_N=0, \quad Y^k_D=d^k. \nonumber
\eea
It is easy to check
that these are compatible with the gauge conditions (\ref{nrorthogauge}).
As was noted before for (\ref{ttbc}), to determine the ranges of $j$ and $k$
(i.e., the number of transverse Dirichlet and Neumann directions) we 
would need
to carry out a supersymmetry analysis; but, given that the partially 
transverse
fivebrane has a finite tension (Eq. (\ref{lt5})), our expectation
for this case is simply that $j=2,\ldots,6$ and $k=7,8,9,10$.
The nonlinearity of the boundary conditions (\ref{ltbc}) (and of the 
equations of motion that follow from (\ref{nromem})) makes it extremely 
difficult to proceed further.
The situation is even worse in the case of a longitudinal M5-brane (OM 
theory), where we expect $Y^l=d^l$, and are then left with the highly 
non-linear requirement that each of the six expressions inside the 
$\delta X^a$ 
and $\delta Y^d$ square brackets
in (\ref{omembdry}) vanish.

\section{Conclusions}
\label{conclsec}

The main aim of this work has been to formulate OM theory in terms of a 
membrane action. Our approach in deriving this action has been to 
directly examine the effect of the OM theory limit (\ref{memlim}) on the 
worlvolume action for an M2-brane, an idea which was pursued for NCOS 
theory by Gomis and Ooguri \cite{go}. Their results for the string case 
served as a general motivation for our work; their method, however, is 
not easily carried over to the membrane case, so we have developed
 an alternative approach.

Our strategy is to employ a Nambu-Goto (as opposed to Polyakov) type 
description for the object in question. As shown in the point-particle 
setting in Section \ref{partsec}, this allows a straightforward 
derivation of the limiting form of the action, and makes the crucial 
cancellation between tensile and `electric' potential energies 
completely transparent. Using this strategy we have reexamined the 
string case in Section \ref{stractsec}, obtaining a 
reparametrization-invariant worldsheet action for Wound string theory, 
Eq.~(\ref{nrngstr}). As explained in \cite{wound,go} and reviewed in the 
Introduction,  besides closed strings and `transverse' D-branes this 
ten-dimensional structure contains the various NCOS theories as states 
associated with `longitudinal' D-branes.  The description of each of 
these objects in terms of the Wound string theory action was studied in 
Section \ref{strspecsec}. For closed strings and transverse D-branes, a 
static gauge choice was shown to reduce the action to free-field form, 
and lead to the expected energy spectra. The static gauge is on the 
other hand incompatible with the boundary conditions relevant to the 
longitudinal D-brane (i.e., NCOS) case.  To obtain a formalism capable 
of describing all cases, it is convenient to work instead in conformal 
gauge. Doing so the action simplifies to the form (\ref{nrpstr}), which 
is nothing but a rewriting of the Gomis-Ooguri action. Even though this 
result is not new, we believe our approach sheds additional light on the 
results of \cite{go}.

In Section \ref{memsec} we turned our attention to the membrane case, 
our main interest. The Nambu-Goto-based treatment was shown in Section 
\ref{memactsec} to yield a reparametrization-invariant action, 
Eq.~(\ref{nrngmem}), for the eleven-dimensional M-theoretic
construct known as 
Wrapped M2-brane (WM2) theory, within which OM theory appears as the 
class of states associated with a `longitudinal' M5-brane 
\cite{wound,go}. 
Just like its string counterpart, 
the closed membrane allows a choice of static gauge, and in Section 
\ref{cmemsec} this was seen to reduce the WM2 action to free-field form. 
Contrary to the standard case, where flat directions in the 
potential give rise to an instability, the WM2 closed membrane spectrum, 
Eq.~(\ref{p0cmem}), was found to be \emph{discrete}, and to possess a 
non-relativistic structure, as expected from the nature of the OM theory 
limit. Under double dimensional reduction it was seen to reproduce the 
closed string spectrum of Wound string theory, just as it should.

Besides closed membranes, WM2 theory contains three qualitatively 
distinct types of M5-branes, the properties of which were analyzed at 
the beginning of Section \ref{omemsec}. It was found in particular that 
`fully transverse' M5-branes are tensionless, an intriguing property 
(noted already in \cite{talk}) which in our opinion merits further 
study. We then proceeded to take some steps towards determining whether 
and how the dynamics of each type of fivebrane is encoded in the WM2 
action appropriate for the corresponding open membrane. For the fully 
transverse M5-brane, it is possible again to work in static gauge, 
thereby obtaining a discrete open membrane spectrum very similar to the 
closed membrane one. Reducing to ten dimensions, this leads to an 
interesting prediction for the excitation spectrum of a transverse 
NS5-brane in Wound IIA string theory (also known to be tensionless 
\cite{talk}). For partially transverse and longitudinal M5-branes, the 
relevant open membrane boundary conditions make it necessary to seek an 
alternative to the static gauge. Use of an `orthonormal' gauge allows 
the WM2 action to be cast in the simplified form (\ref{nromem})  (with 
the additional term (\ref{omextra}) in the OM theory---i.e., 
longitudinal fivebrane--- case). This is then our main result: the 
desired membrane action for WM2/OM theory. Although much simpler than 
the usual Nambu-Goto action, (\ref{nromem}) is still power-counting 
non-renormalizable and leads, as we have noted, to complicated equations 
of motion and non-linear boundary conditions. Under double dimensional 
reduction, it was demonstrated to reduce as expected to the Wound string 
theory action (\ref{nrpstr}).

As in the standard case, this formal reduction
does not necessarily imply that the membrane will dynamically reduce to a string when the 
circle is shrunk to zero size \cite{russo,rt,dwpp}. 
This is closely related
to the possible existence of membrane instabilities.
As we saw in Section \ref{omemsec},  the potential which follows from the
action (\ref{nromem}) has flat directions, which 
leads us to expect a 
continuous spectrum, indicative of multi-particle states.
This seems to be in conflict with our derivation of 
a discrete spectrum, Eq.~(\ref{p0cmem}),
for the closed membrane. As we explained, the origin of the discrepancy is
the fact that the static gauge employed there removes from
the theory the modes responsible
for the instability, and consequently,
the possiblity to have multi-particle states.

Our analysis has been restricted to the bosonic part of the WM2/OM 
theory action, so an important pending task is to work out its 
supersymmetrized version. Knowledge of the full action will be essential 
for a more careful study of the question of nonrenormalizability, and 
more generally, to establish whether the membrane formalism we have 
developed here provides a useful handle on the dynamics of WM2/OM 
theory. 
We should also ask if there exists a regularized version of this membrane
action which facilitates the extraction of physical information.
A successful formulation should, among other things, permit the 
calculation of the fivebrane tensions (\ref{tt5}), (\ref{lt5}), and 
(\ref{ll5}), yield the expected `Newtonian supergravity' interactions 
\cite{go,newt},
and, in the OM theory case, allow a derivation from first 
principles (e.g., through the calculation of $X^M$ correlation 
functions) of the expected open membrane metric and `noncommutativity' 
parameter \cite{bbss1,bbss2,ommetric,ncmem}. Such 
a formulation would hopefully bring us closer to the underlying structure of M 
theory.

\section{Acknowledgements}
AG would like to thank Ulf Danielsson and Mart\'{\i}n Kruczenski for 
collaboration on the issue of the existence of tensionless branes in 
Wound/Wrapped theories, reported on in \cite{talk}. JAG and JDV 
acknowledge partial support from the grants DGAPA-UNAM 
IN117000 and CONACyT-32431-E. The work of AG was supported by 
repatriation grant No.010310 from Mexico's National Council of Science 
and Technology (CONACyT).

\section*{Appendix: Hamiltonian Analysis}

The aim of this appendix is to rederive, by using
the systematics of the Hamiltonian Dirac method, our basic results:
the first order action (\ref{nrpstr})
for  Wound String theory,
and the corresponding result (\ref{nromem})
for the Wound Membrane case.
We will carry out the discussion for the membrane, which includes
the string and the particle as special cases.
Our analysis will be done in a completely gauge-independent
way and we will be able
to recover these results as particular gauge-fixings from a
basic first-order Lagrangian action, up to trivial redefinitions of the
Lagrange multipliers. We will also analyze the role played by the boundary
term (\ref{strbdry}) in this first order theory.

Our starting point is the Nambu-Goto membrane action (\ref{ngmem})
\be \label{ngmem2}
\Sng=-T\int d^3\sigma \left[ \sqrt{-\det g_{\alpha\beta}}
      -\frac{1}{6}C_{NML}\varepsilon^{\alpha\beta\gamma}\p_{\alpha}X^{N}
       \p_{\beta}X^{M}\p_{\gamma}X^{L}\right]~,
\ee
where $g_{\alpha\beta}=g_{MN}\p_{\alpha}X^{M}\p_{\beta}X^{N}$, the world-sheet
space-time labels $\alpha,\beta,
\gamma=0,1,2$, and the space-time labels
  $N,M,L=0,1....10$. 
{}From here
the canonical momenta $P_N$ are given by
\be \label{Pmem}
P_N=\frac{T}{\sqrt{-\det 
g_{\alpha\beta}}}g_{NM}\varepsilon^{\alpha\beta\gamma}
\partial_\alpha X^M (\partial_1 X\cdot\partial_\beta X)
(\partial_2 X\cdot\partial_\gamma X) +{T\over 2} C_{NML}\varepsilon^{rs}
  (\partial_r X^M \partial_s X^L)~,
\ee
where $r,s,t=1,2$ are spatial world-sheet
labels.
As a result of the three-dimensional reparametrization invariance of
(\ref{ngmem2}), the momenta (\ref{Pmem}) conjugate to $X^{N}$ satisfy the
three primary first-class constraints
\bea \label{ngmemconst}\nonumber
\cH_r&\equiv& P_{M}\partial_r X^M~,\qquad r=1,2 \\
\cH&\equiv&\frac{1}{2T}(P_{N}-{T\over 2} C_{NKL}\varepsilon^{rs}
\p_r X^K\p_s X^L)^2  +\frac{T}{2} \det g_{rs}~.
\eea
The first order Lagrangian action associated to the Dirac  total Hamiltonian
is
\be
S=\int d^3\sigma ({\dot X}^N P_N-N\cH-N^r\cH_r)~,
\ee
where $N$ and $N^r$ are arbitrary Lagrange multipliers. This action
can be rewritten in the form
\bea \label{foamem}
&S&=\int d^3\sigma\Bigg\{ P_K ({\dot X}^K +N g^{NK} C_{NML}\varepsilon^{rs}
\p_r X^M\p_s X^L-N^r\p_r X^K)\\
&-&\frac{N}{2}\Bigg[ \frac{g^{NK}P_NP_K}{T}
+ {T\over 4}(g^{IJ}C_{INM}C_{JKL}
+2g_{NK}g_{ML})\varepsilon^{rs}\varepsilon^{tu}
\p_r X^N\p_s X^M
\p_t X^K\p_u X^L\Bigg]\Bigg\}\nonumber~.
\eea
Now it is easy to see the effect
of the limit (\ref{memlim}) in the first order action (\ref{foamem}):
it cancels the quadratic
terms in the longitudinal momenta $P_a$ ($a=0,1,2$), 
rendering the auxiliary fields $P_a$
as new Lagrange multipliers. Indeed, upon taking the limit we have
\bea\label{p0amemh} \nonumber
S&=&\int d^3\sigma\Bigg\{ P_a ({\dot X}^a +
{N\over 2}  \varepsilon^{a}_{bc}\varepsilon^{rs}
\p_r X^{b}\p_s X^{c}-N^r\p_r X^a)
+P_i({\dot Y}^i-N^r\p_r Y^i)\\
& -&\frac{N}{2}\Bigg[ \frac{P_iP_j\delta^{ij}}{T}+
{T\over 2}\varepsilon^{rs}\varepsilon^{tu}
\p_r X\cdot\p_t X(
\p_s Y\cdot \p_u Y
+\lambda\p_s X\cdot \p_u X)
\Bigg]\Bigg\}~,
\eea
where
$i,j,k,l=3...10$ are transverse spatial indices.
The Hamiltonian constraints that follow from this action are
\bea \label{nrngmemconst}
{\cH}_r&\equiv& P_N\p_r X^N~, \\ \nonumber
{\cH}&\equiv&
\frac{\delta^{ij} P_iP_j}{2T}
-{1\over 2}\varepsilon^{a}_{bc}\varepsilon^{rs}
P_a \p_r X^{b}\p_s X^{c}
+\frac{T}{4}\varepsilon^{rs}\varepsilon^{tu}
\p_r X\cdot\p_t X(
\p_s Y\cdot \p_u Y
+\lambda\p_s X\cdot \p_u X)
\eea
which can be recognized as the limiting form of the original Hamiltonian
constraints (\ref{ngmemconst})
associated with
the action (\ref{ngmem2}). The structure
of the Hamiltonian constraint ${\cH}$ makes
the non-relativistic nature of the theory apparent (linear in $P_a$).

We can now eliminate the auxiliary variables $P_i$ using their equations
of motion, to obtain the Lagrangian action
\bea\label{p0amem} \nonumber
S&=&\int d^3\sigma\Bigg\{ P_a ({\dot X}^a +
{N\over 2}  \varepsilon^{a}_{bc}\varepsilon^{rs}
\p_r X^{b}\p_s X^{c}-N^r\p_r X^a)
+\frac{T}{2N}({\dot Y}^i-N^r\p_r Y^i)^2\\
& -&\frac{N}{4}\Bigg[
T\varepsilon^{rs}\varepsilon^{tu}
\p_r X\cdot\p_t X(
\p_s Y\cdot \p_u Y
+\lambda\p_s X\cdot \p_u X)
\Bigg]\Bigg\}~,
\eea
where the variables $P_a$ are seen to play the role of Lagrange multipliers
enforcing the Lagrangian constraint
\be\label{lagconst}
{\dot X}^a +
{N\over 2}  \varepsilon^{a}_{bc}\varepsilon^{rs}
\p_r X^{b}\p_s X^{c}-N^r\p_r X^a=0~.
\ee
The Lagrangian action (\ref{p0amem}) is the main result of this Appendix.
As we will see below, the actions (\ref{nromem}) and (\ref{nrpstr}) can be recovered
as particular gauge-fixings of this general result.

Variation of the Hamiltonian action (\ref{p0amem}) 
leads to the boundary term
\be
n_r\frac{\partial {\cal L}}{\partial_r X^N}\delta X^N=0~,
\ee
where $n_r$ is a unit vector normal to the boundary 
surface.
There are several cases to consider, depending on whether
the membrane is closed or open, and if open, whether it ends on a
longitudinal, partially transverse, or fully transverse five-brane.

The above boundary term can be expanded into
\bea\label{canonicalbound}
\nonumber
&n_r&\Bigg[(N\varepsilon^a_{bc}
\varepsilon^{rs}P_a \p_sX^c
-{T\over 2}N\eta_{ab}
\varepsilon^{rs}\varepsilon^{tu}\p_tX^a(
  \p_s Y\cdot\p_u Y-2\lambda\p_sX\cdot\p_uX) - N^r P_b)~\delta X^b\\
&+&\left({T\over 2}N\delta_{ij}\varepsilon^{rs} \varepsilon^{tu}\p_tY^j
\p_sX\cdot\p_uX -N^r P_i\right)~\delta Y^i\Bigg]=0~.
\eea
It is a difficult task to solve the equations of
motion with these boundary
conditions for any gauge. 
We therefore fix a particular gauge compatible
with the closed, transverse, longitudinal or mixed cases. 

In the closed membrane 
case we have only the periodicty requirement, which allow
the static gauge choice
 $X^0=c\sigma^0$, $X^1=2w_1 R_1\sigma^1$, $X^2=2w_2
R_2\sigma^2$. This gauge
is canonical and implies $N=c/4w_1w_2R_1R_2$
and $N^r=0$. These fixed Lagrangian multipliers and the static gauge
solve (\ref{lagconst}). The reduced action can be obtained by
implementing this gauge in the action (\ref{p0amem}), and the result is
the non-relativistic membrane given by (\ref{nrsmem}).

The corresponding analysis for the  open membrane case
 can be simplified in the orthonormal gauge.
This gauge is noncanonical, and is implemented by imposing the conditions
$N=1$ and
$N^r=0$ on the Lagrange multipliers.
A general property of a constrained
system is that the Lagrange multipliers associated with primary
first class constraints can always be obtained as Lagrangian functions using
the equations of motion for the auxiliary fields $P_N$,
\be
{\dot X}^K +{1\over 2}N g^{NK} C_{NML}\varepsilon^{rs}
\p_r X^M\p_s X^L-N^r\p_r X^K-N \frac{g^{NK}P_N}{T}=0,
\ee
the constraints and the definition of the momenta (\ref{Pmem}).
This procedure is the inverse
of the Legendre transformation in the extended phase space where the
Lagrangian multipliers are promoted to dynamical variables. In our case
\be
N^r\p_rX^N\p_1X^K g_{NK}=\p_1X^N{\dot X}^K g_{NK},\qquad
N^r\p_rX^N\p_2X^K g_{NK}=\p_2X^N{\dot X}^K g_{NK}
\ee
and
\be
N=\frac{\sqrt{-\det g_{\alpha\beta}}}{\det g_{rs}}.
\ee
By implementing the noncanonical gauge
$N=1/L, N^r=0$ on these relations we obtain the corresponding
relations (\ref{orthogauge}) of the main text.

When evaluated in the appropriate gauge, the results
in this Appendix reduce to the ones presented in the main text.
In particular, up to a trivial redefinition of the
Lagrange multipliers, the first order action (\ref{nromem}) can be obtained
by enforcing the above noncanonical gauge in the action
(\ref{p0amem}). The corresponding boundary terms are gauge-fixed
versions
of the general phase-space boundaries given by (\ref{canonicalbound}).

\end{document}